\documentclass{article}
\usepackage[utf8]{inputenc}

\usepackage{amsmath, amsthm, amssymb,color,fullpage,url,booktabs}  
\usepackage{hyperref}
\usepackage{graphicx}
\usepackage{cleveref}
\usepackage{tikz}
\usepackage{authblk}
\usepackage[caption=false]{subfig}

\title{Enhanced Framework of Quantum Approximate Optimization Algorithm and Its Parameter Setting Strategy}
\author[1]{Mingyou Wu}
\author[1,2]{Zhihao Liu}
\author[1,2]{Hanwu Chen}
\affil[1]{\small School of Computer Science and Engineering, Southeast University, Nanjing 211189, China}
\affil[2]{\small Key Laboratory of Computer Network and Information Integration (Southeast University), Ministry of Education, Nanjing, 211189, China}

\begin{document}

\maketitle
\begin{abstract}
An enhanced framework of quantum approximate optimization algorithm (QAOA) is introduced and the parameter setting strategies are analyzed. The enhanced QAOA  is as effective as the QAOA but exhibits greater computing power and flexibility, and with proper parameters, it can arrive at the optimal solution faster. Moreover, based on the analysis of this framework, strategies are provided to select the parameter at a cost of $O\left(1\right)$.  Simulations are conducted on randomly generated 3-satisfiability (3-SAT) of scale of 20 qubits and the optimal solution can be found with a high probability in iterations much less than $O(\sqrt{N})$
\end{abstract}

\section {Introduction}
In 2014, Farhi \emph{et al}. \cite{Farhi2014} proposed a quantum classical hybrid variational method, the quantum approximate optimization algorithm (QAOA) for combinatorial optimization problems (COP). The QAOA defines the problem Hamiltonian $H_C$ and use the transverse field as the mix Hamiltonian $H_B$. These two Hamiltonians are alternately applied to the quantum state and the evolution time $\gamma$ and $\beta$ are introduced as parameters. With proper parameters, the expectation of the quantum state gradually grows and after enough iterations, the optimal solution will be obtained. 

Recently, Farhi \emph{et al}. \cite{Farhi2020} pointed out that the QAOA need to see the whole graph and used the QAOA+ as an example. The QAOA+ adopts the information of graph structure into the problem Hamiltonian instead of preparing a specific initial state. Based on the comparison between QAOA and QAOA+ on maximum independent set (MIS), an enhanced framework of QAOA is proposed, which is effective and more flexible. Besides, the complexity of the enhanced QAOA is analyzed and a parameter initial strategy of costs of $O(1)$ is provided. This strategy naturally applies to 3-satisfiability (3SAT) and in a single simulation the satisfiability can be determined with a probability around 50\% for $O(n\log{m})$ iterations, where $m$ is the number of constraints valued in $O(n^3)$. For general NP-complete problem, a reduction to 3-SAT is always available, and strategies are advised to adjust the parameters.

The reminder of this paper is o rganized as follows. Section 2 briefly reviews the QAOA and QAOA+ for MIS. In Section 3, a basis is introduced and developed on this, the enhanced QAOA is proposed and analyzed. Section 4 presents the practical meaning and the setting strategies of parameters, and corresponding simulations are conducted. Finally, Section 5 concludes this paper.

\section{Review of the QAOA }
For a combinatorial optimization problem of scale $n$ with $m$ clauses, the objective function can be expressed as follows:
\begin{equation}\label{eq1}
	C\left( z \right)=\sum\limits_{\alpha =1}^{m}{{{C}_{\alpha }}\left( z \right)},
\end{equation}
where string $z\in {{\left\{ 0,1 \right\}}^{n}}$ and ${{C}_{\alpha}}$ stands for a certain clause $\alpha$,
\begin{equation}\label{eq2}
	{{C}_{\alpha }}=\left\{ \begin{matrix}
	\text{1 if }z\text{ satisfies the clause }\alpha,   \\
	\text{0 if }z\text{ does not satisfy.}  \\
\end{matrix} \right.
\end{equation}

COP asks for a string $z$ that maximizes the objective function $C(z)$. And the approximate optimization demands a string $z$ for which $C(z)$ is close to the optimal of $C$.

In the QAOA, the initial state is usually the equal superposition state
\begin{equation}\label{eq3}
	\left| s \right\rangle =\frac{1}{\sqrt{{{2}^{n}}}}\sum\limits_{j}{\left| j \right\rangle }.
\end{equation}
And the evolution operators of QAOA is a series of operators interleaved with 
${{U}_{C}}\left( \gamma  \right)={{e}^{-i\gamma {{H}_{C}}}}$ and ${{U}_{B}}\left( \gamma  \right)={{e}^{-i\beta {{H}_{B}}}}$. $\gamma$ and $\beta$ are parameters between 0 and $2\pi$ and varies during the procedure of QAOA. $H_C$ is the problem Hamiltonian which is determined by the problem as
\begin{equation}\label{eq4}
	{{H}_{C}}=diag\left\{ C\left( {{z}_{i}} \right) \right\},
\end{equation}
and $H_B$ is the mix Hamiltonian, which usually is the transverse field written as
\begin{equation}\label{eq5}
	{{H}_{B}}=\sum\limits_{j}{\sigma _{j}^{x}}.
\end{equation}
For $p$ iterations the system state is
\begin{equation}\label{eq6}
	\left| {{z}_{p}} \right\rangle ={{U}_{B}}\left( {{\beta }_{p}} \right){{U}_{C}}\left( {{\gamma }_{p}} \right)\ldots {{U}_{B}}\left( {{\beta }_{1}} \right){{U}_{C}}\left( {{\gamma }_{1}} \right)\left| s \right\rangle ,
\end{equation}
and here $p$ is call the depth of QAOA. $\left| {{z}_{p}} \right\rangle$ can also be denoted as $\left| \gamma ,\beta  \right\rangle$, and the expectation of state $\left| {{z}_{p}} \right\rangle $ on $H_C$ is 
\begin{equation}\label{eq7}
	{{F}_{p}}\left( \gamma ,\beta  \right)=\left\langle  \gamma ,\beta  \right|H_C\left| \gamma ,\beta  \right\rangle.
\end{equation}
${{F}_{p}}\left( \gamma ,\beta  \right)$ can be used to select the parameters such that ${{F}_{p-1}}\left( \gamma ,\beta  \right)<{{F}_{p}}\left( \gamma ,\beta  \right)$.

The maximum independent set (MIS) problem asks for the independent set of the largest possible size for the given graph. Farhi \cite{Farhi2014} first applies the QAOA to the MIS, and use 
\begin{equation}\label{eq8}
	C\left( z \right)=\sum\nolimits_{j}{{{z}_{j}}}
\end{equation}
as the objective function. This definition only considers the number of vertices in each solution, so a specialized initial state is required, and the quantum adiabatic algorithm is applied to prepare the superposition of all feasible solutions. Recently Farhi \cite{Farhi2020} suggested that the QAOA should consider the whole graph, and put forward the QAOA+ with a new objective function
\begin{equation}\label{eq9}
	{{C}_{+}}\left( z \right)=\sum\nolimits_{j}{{{z}_{j}}}-\sum\limits_{u,v}{{{A}_{u,v}}{{z}_{u}}{{z}_{v}}},
\end{equation}
where ${{A}_{u,v}}$ is the element of the adjacent matrix of given graph. This objective function considers the basic structure of the graph and has a more powerful computing power, which can handle the MIS problem without the help of an appended initial state preparation.

\section{The enhanced framework of QAOA}
Considering the problem Hamiltonians of QAOA and QAOA+ for MIS, the detailed expression is
\begin{equation}\label{eq10}
	\left\{ \begin{matrix}
		& {{H}_{C}}=\sum\limits_{k}{{{P}_{k}}}, \\ 
		& {{H}_{C+}}=\sum\limits_{k}{{{P}_{k}}}-\sum\limits_{u,v}{{{A}_{u,v}}{{P}_{u,v}}}, \\ 
	\end{matrix} \right.
\end{equation}
where $P_k$ denotes operator $P=diag\left\{0,1\right\}$ on the $k$-th qubit, $P_{u,v}$ denotes $P$ on the $u$-th and $v$-th qubits, and $A_{u,v}$ is the element of the adjacent matrix of graph. Both the problem Hamiltonians can be expressed as a linear combination of projection operators. Expand these projection operators to 
\begin{equation}\label{eq11}
	{{p}_{j}}=\otimes _{i=1}^{n}{{\left( {{P}_{i}} \right)}^{{{j}_{i}}}}={{\left( {{P}_{1}} \right)}^{{{j}_{i}}}}\otimes {{\left( {{P}_{2}} \right)}^{{{j}_{2}}}}\otimes \ldots \otimes {{\left( {{P}_{n}} \right)}^{{{j}_{n}}}}.
\end{equation}
Let $Z_i$ denote $Z$ on the $i$-th qubit. The Walsh operator \cite{Welsh2014} on $n$ qubits is
\begin{equation}\label{eq12}
	{{w}_{j}}=\otimes _{i=1}^{n}{{\left( {{Z}_{i}} \right)}^{{{j}_{i}}}}={{\left( {{Z}_{1}} \right)}^{{{j}_{i}}}}\otimes {{\left( {{Z}_{2}} \right)}^{{{j}_{2}}}}\otimes \ldots \otimes {{\left( {{Z}_{n}} \right)}^{{{j}_{n}}}}.
\end{equation}
Actually, $Z=I-2×P_i$, and ${w_j}$ can be represented by ${p_j}$. ${w_j}$ is an orthonormal basis of dialog matrixes of dimension $2^n$ and $e^{-i\theta_j w_j}$ can be implemented by $(n)$ basic gates \cite{Welsh2014}. Obviously, ${p_j}$ also consists a basis and the implementation cost of $e^{-i\gamma_j p_j}$ is $O(n)$ \cite{SCP2018}.

Therefore, the unitary operator of any problem Hamiltonian can be rewritten as
\begin{equation}\label{eq13}
	{{U}_{C}}\left( \gamma  \right)={{e}^{-i\sum\nolimits_{j=0}^{N-1}{\gamma {{p}_{j}}}}}.
\end{equation}
Replacing $\gamma$ in Eq. (\ref{eq13}) with $\gamma_j$, a new unitary operator can be written as
\begin{equation}\label{eq14}
	{{U}_{Ce}}\left( \gamma  \right)={{e}^{-i\sum\nolimits_{j=0}^{N-1}{{{\gamma }_{j}}{{p}_{j}}}}}.
\end{equation}
This is the original idea and basic form of the enhanced QAOA. The evolution operator $e^{-i\gamma H_C}$ is replaced by a sequence of control evolution gates that
\begin{equation}\label{eq15}
	C{{R}_{j}}\left( \gamma  \right)={{e}^{-i{{\gamma }_{j}}{{p}_{j}}}},
\end{equation}
where if the $x$-th qubit is a control qubit, then $j_x=1$. In fact, for the Hamiltonian of majority of COP only a few bases in $\left\{p_j\right\}$ are used. Therefore, the layer of QAOA is defined as the maximum of $d(j)$ of all control evolution gates, where 
\begin{equation}\label{eq16}
	d\left( j \right)=\sum\limits_{x=1}^{n}{{{j}_{x}}}
\end{equation}
is the number of control bits. 

Denote COP with constraints that engage no more than $k$ variables as $COP_k$. Without an extern computing power or extra information, a $k$-layer QAOA can solve $COP_k$ but cannot deal with $COP_{k+1}$. MIS after specific initial state preparation is in $COP_1$. In fact, the evolution operators of 1-layer QAOA are local on a single qubit and the entanglement is invariable. It means the 1-layer QAOA does not provide any computation power, but only present the computation basis closest to the optimal solution. NP-complete problem and $COP_2$ can reduce to each other in polynomial time. In fact, max-cut is in $COP_2$ and every problem in $COP_2$ can be reduced as a max-cut problem on a weighted graph with loop. $COP_k$ is the NP-optimization problem for a constant $k$ such as MIS, 3SAT and E3Lin2 \cite{Farhi2015}. $COP_n$ is the hardest COP of scale $n$. In classical computer, it cost exponentially to evaluate the quality of a solution, and in quantum computer, the cost of the implementation of the $U_C$ is also exponential. 

It is clear that the enhanced QAOA has a similar framework to QAOA with the same implementation cost, and the parameter $\pmb{\gamma} =\left( {{\gamma }_{j}} \right)$ enables the Hamiltonian to vary in a larger space which would result in an increase of the computation capability. For example, consider a simple comparison between the standard QAOA and the enhanced QAOA of 1-layer for the MIS, and the $H_C$ can be respectively written as
\begin{equation}\label{eq17}
	\left\{ \begin{matrix}
		{{H}_{Cs}}=\sum\limits_{j}{{{P}_{j}}},  \\
		{{H}_{Cg}}=\sum\limits_{j}{{{\gamma }_{j}}{{P}_{j}}}.  \\
	\end{matrix} \right.
\end{equation}
Because of the lack of control evolution gates that $d(l)=2$, the 1-layer standard QAOA of $n$ qubits cannot represent majority of the constraints and is unable to deal with MIS of scale $n$. As for the enhanced QAOA, with specific selected parameters $\pmb{\gamma}$, the enhanced QAOA can solve MIS by increasing the parameters of the bases that contain vertices in maximum independent set and decreasing the others. For this case, the computation capability of the 1-layer enhanced QAOA mainly comes from the parameters setting, i. e., the parameters optimizer, and the correctness mainly depends on the optimizer. The interface of the enhanced QAOA available for classical computing power increases and so is the computation capability. But when applying the enhanced QAOA to certain problem, the quantum computation capability ought to be the main component to execute the calculation task, and the classical computer provides assistance, so the layer should be at least as the same as the standard QAOA.

For a fixed layer, the enhanced QAOA can arrive at the target state faster than the standard QAOA. Obviously, the enhanced QAOA cannot be slower than the standard QAOA. For convenience, the parameters $\pmb{\gamma}$ are decomposed into two parts as $\pmb{\gamma}=\gamma_s \pmb{\gamma_r}$, where $\gamma_s$ is the global phase, and $\pmb{\gamma_r}$ is the relative phase. The parameters $\pmb{\gamma_r}$ of the standard QAOA are determined by the constraints and are static during the evolution of algorithm. Use contradiction and suppose the enhanced QAOA is as fast as the standard QAOA. When optimizing the expectation $F\left(\gamma_s,\pmb{\gamma_r}\right)$, $F_m\left(\gamma_s\right)$ should be equal to $F_m\left(\gamma_s,\pmb{\gamma_r}\right)$, that is
\begin{equation}\label{eq18}
	\frac{\partial F({{\gamma }_{s}},\pmb{{\gamma }_{r}})}{\partial \pmb{{\gamma }_{r}}}=\frac{\partial F({{\gamma }_{s}})}{\partial \pmb{{\gamma }_{r}}}=0,
\end{equation}
that is, $\pmb{\gamma_r}$ is independent to $F(\gamma_s,\pmb{\gamma_r})$. This is obviously wrong, such as the 1-layer standard and enhanced QAOA on MIS, and the latter can arrive a larger expectation.

\section{Parameter setting and simulation result}
Consider the normalization of $\pmb{\gamma_r}$. Firstly, the Grover Hamiltonian $diag\left\{1,0,…,0\right\}$ \cite{PRL1997} applied as a search of QAOA \cite{PRA2017} should be normalized and so is the multi-solutions case. And with $\gamma_s=\pi$, $U(H_C,\pmb{\gamma})$ can best distinguish the optimal and non-optimal solutions. For general case, the best normalization is linearly mapping the goal values of solutions from $[C_{min},C_{max}]$ to $[0,1]$. However, $C_{max}$ cannot be directly normalized to 1 because the value of $C_{max}$ is the algorithm target. Instead, ${{C}_{lim}}=\sum\nolimits_{j}{{{\gamma }_{r,j}}}$ is adopted and ${{C}_{lim}}\ge {{C}_{max}}$. Therefore, $\sum\nolimits_{j}{{{\gamma }_{r,j}}}$ should be normalized to 1 and when negative constraints are occupied, $\sum\nolimits_{j}{\left| {{\gamma }_{r,j}} \right|}=1$.

\begin{figure}
	\begin{minipage}[t]{0.5\linewidth}
		\centering
		\subfloat[Case with 2M constraints] {\label{fig1:a}\includegraphics[height=6cm,width=8cm]{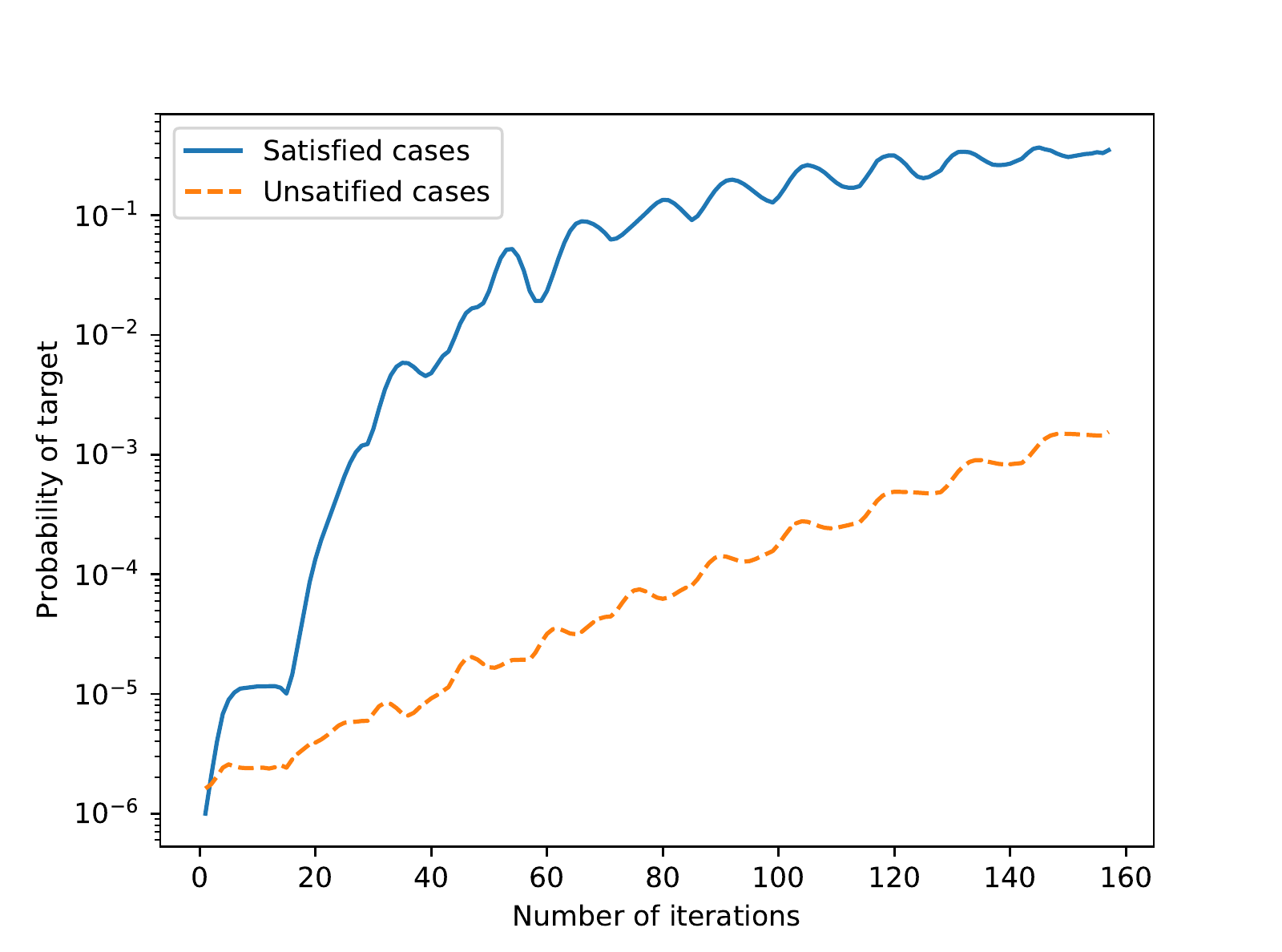}}
	\end{minipage}\begin{minipage}[t]{0.5\linewidth}
		\centering
		\subfloat[Case with 4M constraints] {\label{fig2:b}\includegraphics[height=6cm,width=8cm]{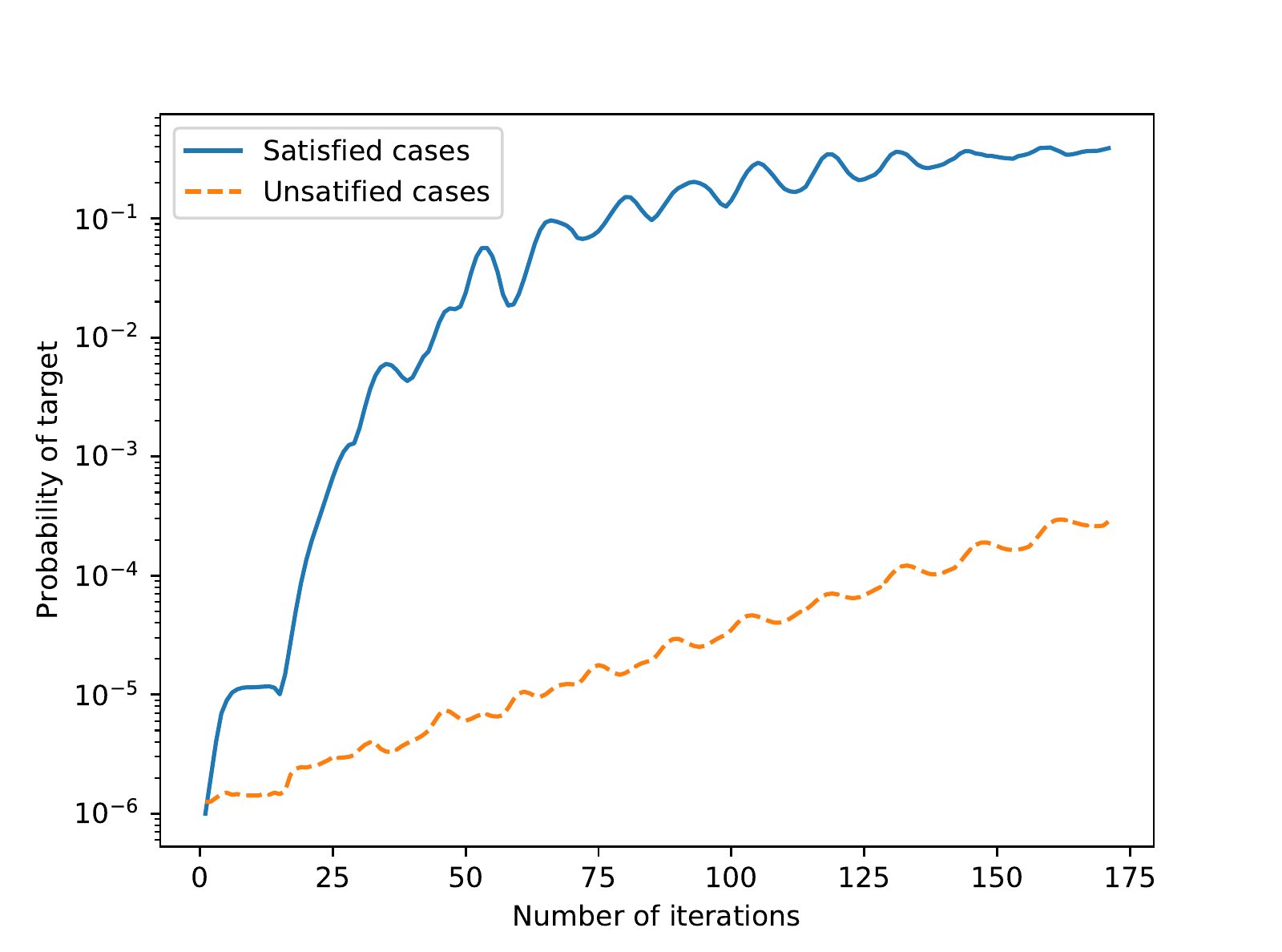}}
	\end{minipage}
	\caption{The average probability of the target computational basis during iterations.}\label{fig1}
\end{figure}

\begin{figure}
	\begin{minipage}[t]{0.5\linewidth}
		\centering
		\subfloat[Case with 0.5M constraints]{\label{fig2_0.5M}\includegraphics[height=5.4cm,width=7cm]{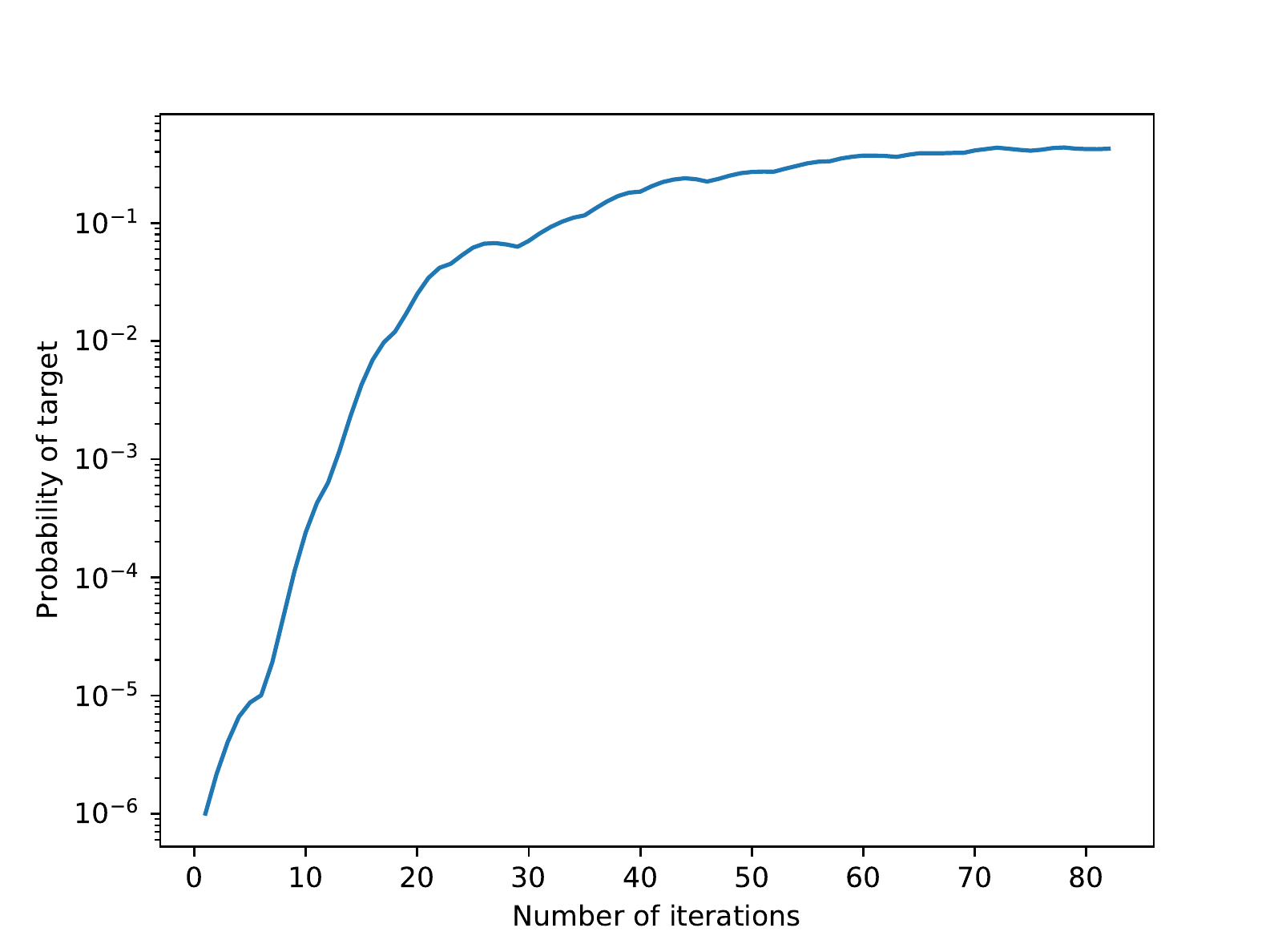}}
	\end{minipage}\begin{minipage}[t]{0.5\linewidth}
		\centering
		\subfloat[Case with M constraints]{\label{fig2_1M}\includegraphics[height=5.4cm,width=7cm]{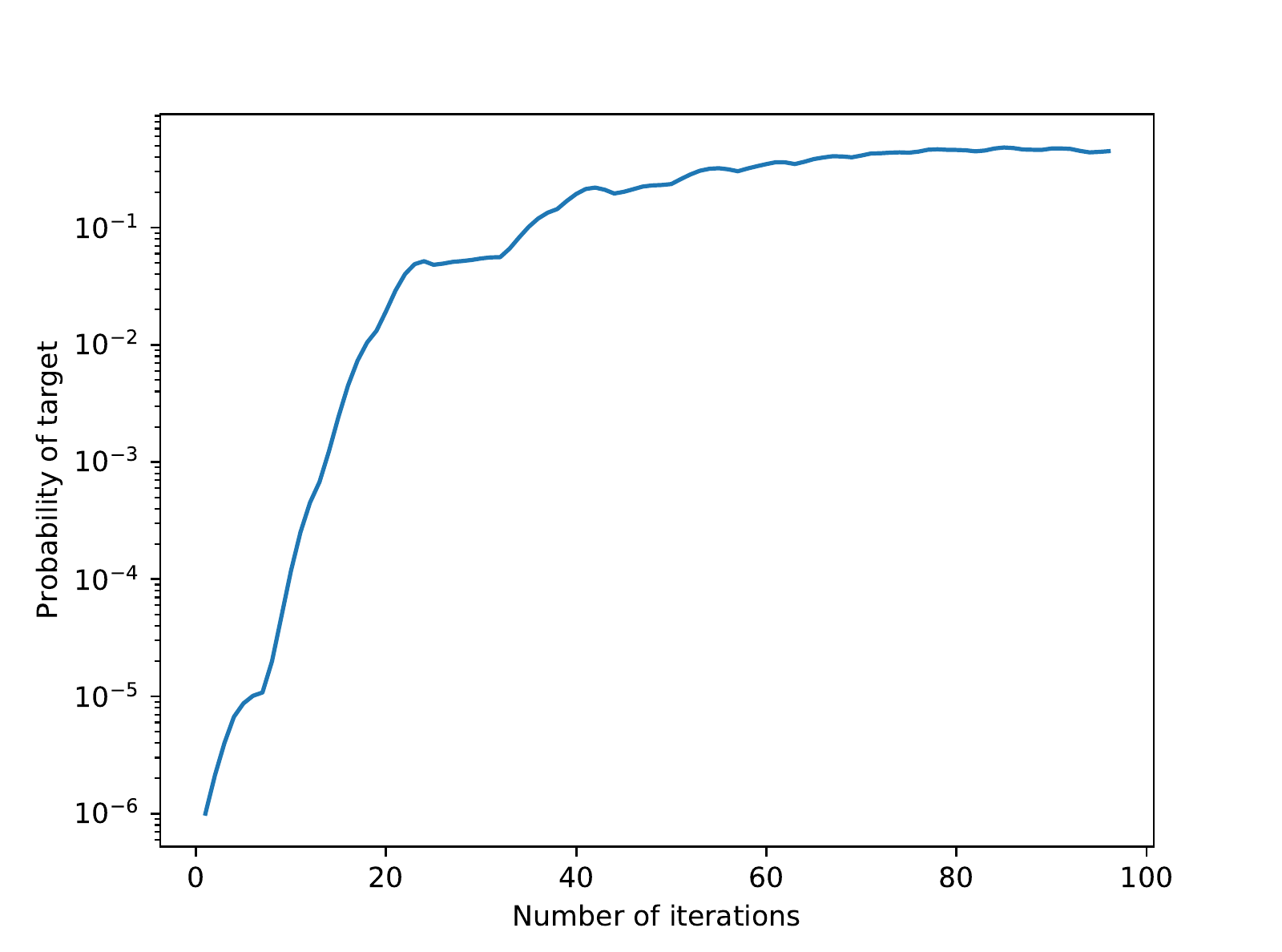}}
	\end{minipage}
	\begin{minipage}[t]{0.5\linewidth}
		\centering
		\subfloat[Case with 2M constraints]{\label{fig2_2M}\includegraphics[height=5.4cm,width=7cm]{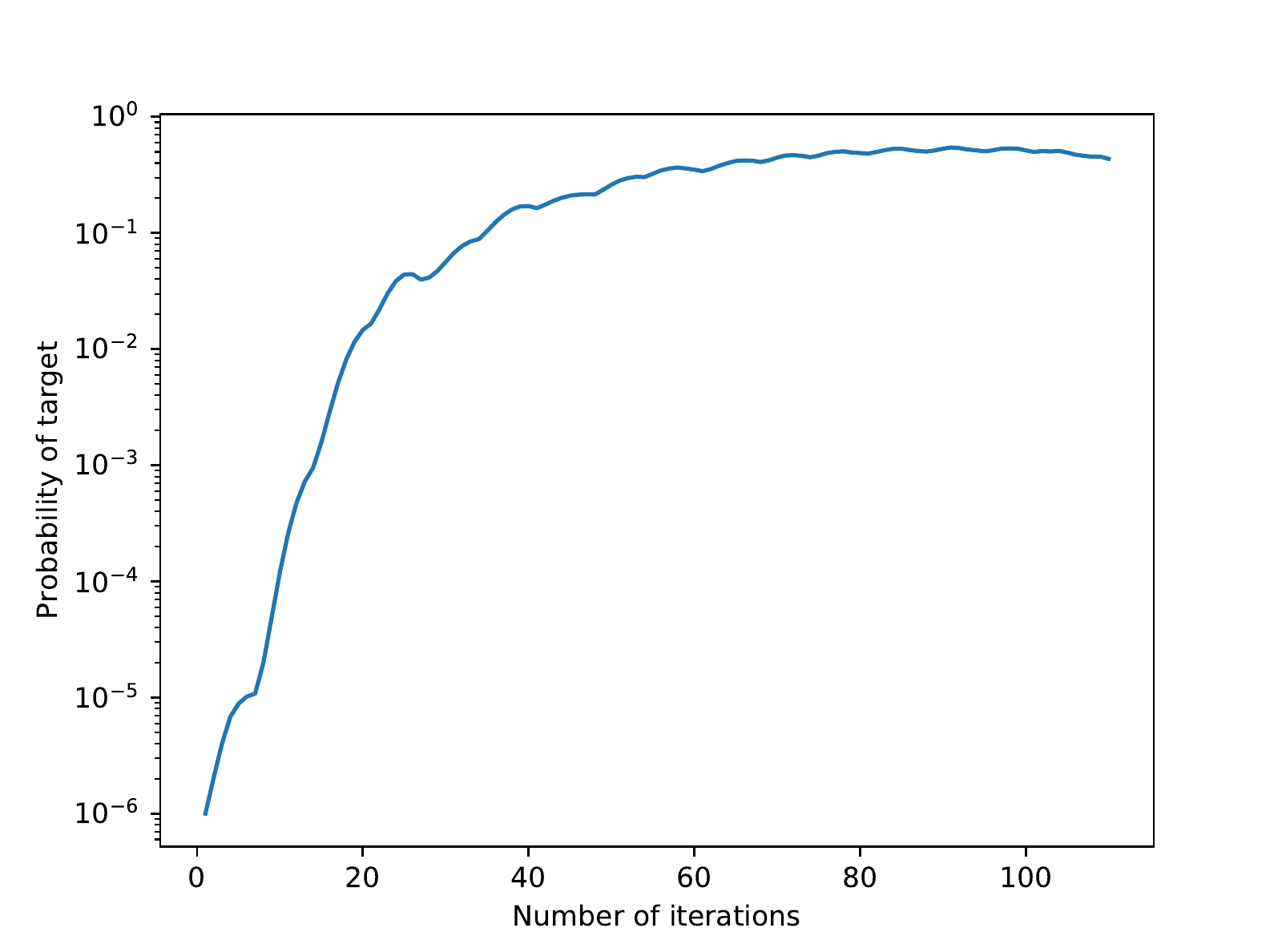}}
	\end{minipage}\begin{minipage}[t]{0.5\linewidth}
		\centering
		\subfloat[Case with 4M constraints]{\label{fig2_4M}\includegraphics[height=5.4cm,width=7cm]{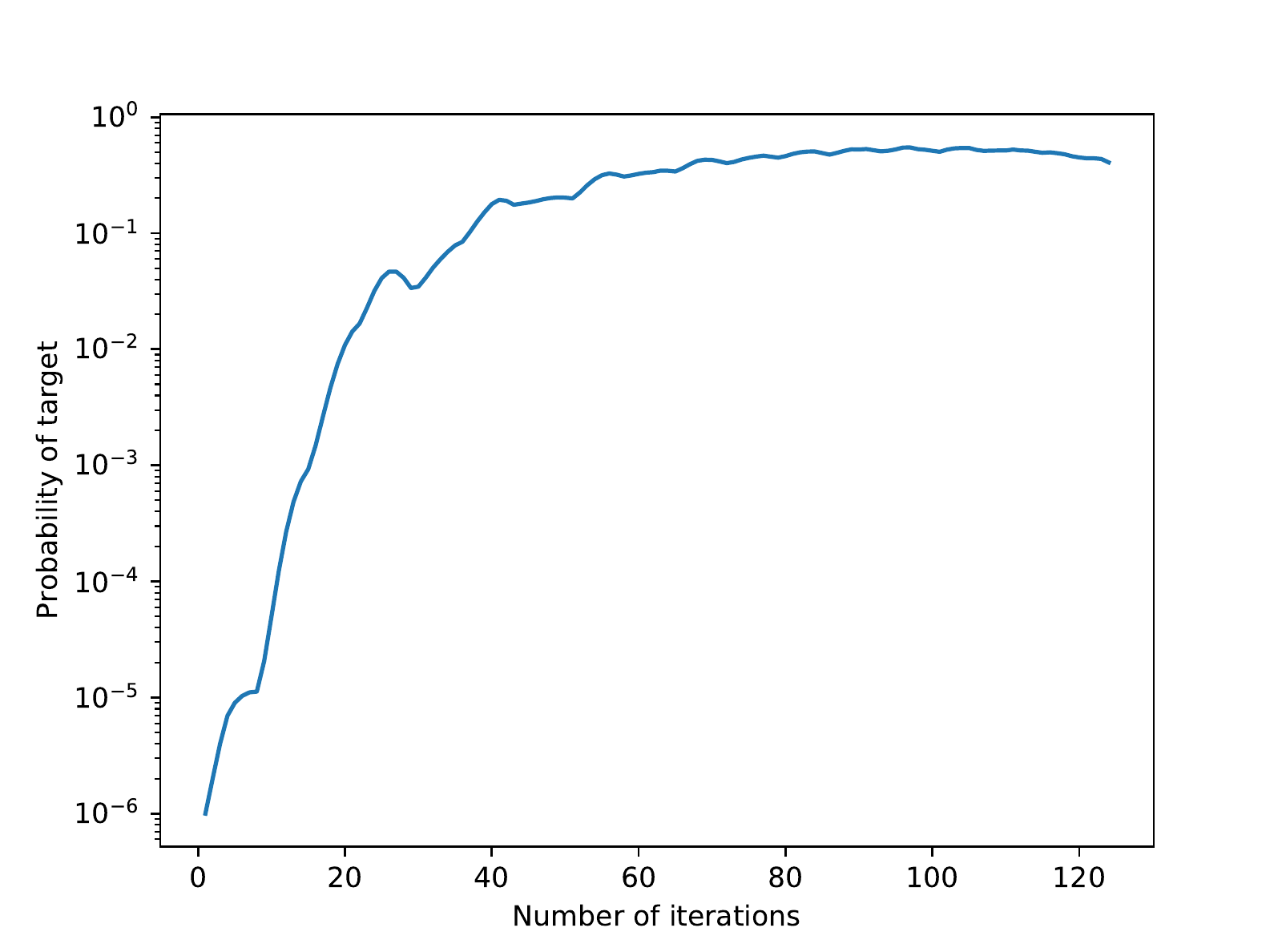}}
	\end{minipage}
	\caption{The average probability of the target computational basis for a decreased iterations.}\label{fig2}
\end{figure}

This normalization strategy naturally applies to satisfiability problem. Using 3-SAT as example, problem is satisfiable if $C_{lim}=C_{max}$, that is, the maximum of the eigenvalue of normalized Hamiltonian is 1. Noting the periodicity of $e^{-i\theta}$, with parameter $\gamma_s=(2t+1)\pi$ and any $t\in z$, the effective phase shift of optimal solution is always $\pi$, but that of the non-optimal solution varies with the change of $t$. By modifying $t$, the angle between the optimal and non-optimal solution would increase. As for unsatisfiable case, the larger the difference between $C_{lim}$ and $C_{max}$, the smaller the probability to find the solution of $C_{max}$. The CTQW part actually has the form of 
\begin{equation}\label{eq19}
	{{e}^{-i\beta {{H}_{B}}}}=H{{e}^{-i\beta {{H}_{Z}}}}H,
\end{equation}
where
\begin{equation}\label{eq20}
	{{H}_{Z}}=\sum\limits_{m=0}^{n-1}{{{Z}_{m}}}
\end{equation}
and $Z_m$ is $Z$ on the $m$-th qubit. Therefore, $1/n$ is a good parameter for $\beta$ because the eigenvalue of $H_Z$ can be normalized to $[-1,1]$. As for $\gamma$, a linear increase parameter is applied. Here the number of iterations is set to be ${n\log m}/{\sqrt{2}}\;$ and the maximum of $t$ is $\log m $, namely, for $p$-depth QAOA, 
\begin{equation}\label{eq21}
	\gamma =\left( \left\lfloor \frac{2\sqrt{2}p}{n} \right\rfloor +1 \right)\pi,
\end{equation}
where $1\le p \le {n\log m}/{\sqrt{2}}\;$. The simulation results of scale of 20 qubits on randomly generated instance of 3-SAT are shown in Figure \ref{fig1} and Table \ref{tab1}. For satisfied case, firstly the solution as randomly selected and then the constraints satisfied by the solution are randomly generated. And for unsatisfied case, the constraints are randomly generated and the satisfied instances are excluded. The number of constraints values $2M$ and $4M$ and both satisfied and unsatisfied cases are considered, where $M=C_{20}^3$. The number of repeated experiments is 50 times for each case. In fact, a fewer number iterations is also feasible as $n\log \left( 2m/n \right)/\sqrt{2}$ and the maximum of $t$ is $\log \left( 2mn \right)$. The simulation results are shown in Figure \ref{fig2} and Table \ref{tab2}. The number of constraints values $0.5M$, $M$, $2M$ and $4M$ and only satisfied case are presented. 

\begin{table}  
	\centering
	\caption{The average of the maximum probability of target during iterations.} \label{tab1}
	\begin{tabular*}{13.5cm}{ccccc}  
		\hline  
		constraints & $2M$\&satisfied  & $2M$\&unsatisfied & $4M$\&satisfied & $4M$\&unsatisfied \\  
		\hline  
		max probability  & 0.4392 & 0.0018 & 0.4747 & 0.0003 \\  
		\hline  
	\end{tabular*}  
\end{table}  

\begin{table}  
	\centering
	\caption{The average of the maximum probability for a decreased iterations.} \label{tab2}
	\begin{tabular*}{8.6cm}{ccccc}  
		\hline  
		constraints & $0.5M$  & $M$ & $2M$ & $4M$ \\  
		\hline  
		max probability  & 0.4815 & 0.5367 & 0.5859 & 0.5921 \\  
		\hline  
	\end{tabular*}  
\end{table}  

However, for general COP, a satisfiable solution is very rare, this is, $C_{max}$ is generally much smaller than $C_{lim}$. Noting the difference ${{\Delta }_{C}}=\left| {C_{lim}}-{{C}_{max}} \right|$, with the growth of the scale of problem, the depth required to arrive the optimal solution increases, and the influence of $\Delta_{C}$ on optimization might become greater and unpredictable. Therefore, specific strategies are required to adjust the Hamiltonian. In fact, the range of the solution of a specific COP can be obtained by probability theory and combinatorial Mathematics, and can be adopted as prior knowledge. By multiplying a specific factor which can be gradually adjusted, the maximum of the eigenvalue of problem Hamiltonian can be approximately normalized to 1. Besides, the measurement result ${e_j}\left( z \right)={\left\langle  z \right|{{p}_{j}}\left| z \right\rangle }\;$ of repeat experiments presents the importance of $p_j$ and can be used to adjust the parameters $\gamma_{r,j}$. And the normalized $\left\{ e_{j}^{\alpha }{{\gamma }_{r,j}} \right\}$ can be adopted as the parameters of next experiment, where $\alpha$ is the adjusting factor and $\alpha \ge 0$. 


\section {Discussion and conclusion}
The enhanced QAOA introduced in this paper inherits the properties of the QAOA without any extra cost, and moreover, exhibits many superiorities. The parameters $\gamma$ and $\beta$ of the standard QAOA is actually the evolution time and the problem Hamiltonian is static during the iterations. However, with the additional parameters $\pmb{\gamma_r}$, the enhanced QAOA can adjust the problem Hamiltonian during algorithm process, which can reduce the complexity and provides interface for classical computing power. Furthermore, the extra computation capability is adjustable and offers more options for researchers. This paper defines the layer of the QAOA that determines the upper bound of the implementation complexity, and presents a series of problems ${COP}_k$ that reflect the upper bound of the computability of the QAOA of certain layer. Meanwhile, the QAOA of different layers also provides reference models for the corresponding problems. QAOA provides a scheme combining quantum and classical computing power, while the enhanced QAOA presents a new view for the architecture of the QAOA, and would be useful to reconsider and organize the previous work. 

This enhanced framework of the QAOA more clearly shows the piratical meaning of the parameters and based on this, parameter setting strategies of the enhanced QAOA are proposed. The simulation shows its efficiency on 3-SAT, but limited by the simulation complexity of quantum system, only cases with 20 qubits are analysis. Further experimental and theoretical analysis are urgently required. Besides, the enhanced QAOA reveals other issues like the analysis of the alteration of Hamiltonian under certain parameter setting strategy. Nevertheless, the enhanced QAOA does not show advantage on matters like the depth analysis of QAOA, the standard QAOA is still needed in some theoretical derivation.

\end{document}